\documentclass[runningheads]{llncs}
\usepackage{graphicx}
\usepackage{setspace}
\usepackage{amsfonts}
\begin{document}
\title{Secrecy-Verifiability Paradox in Smart Contracts}
%
%
\author{Nguyen Ha Thanh}
\authorrunning{Nguyen Ha Thanh}
%
\institute{National Institute of Informatics, Tokyo, Japan}
\maketitle              
\begin{abstract}
The trade-off of secrecy is the difficulty of verification. This trade-off means that contracts must be kept private, yet their compliance needs to be verified, which we call the \textit{secrecy-verifiability paradox}.
However, the existing smart contracts are not designed to provide secrecy in this context without sacrificing verifiability.
Without a trusted third party for notarization, the protocol for the verification of smart contracts has to be built on cryptographic primitives.
We propose a blockchain-based solution that overcomes this challenge by storing the verifiable evidence as accessible data on a blockchain in an appropriate manner. This solution allows for cryptographic data verification but not revealing the data itself. In addition, with our proposal, it is possible to verify contracts whose form of existence has been destroyed as long as the contract is real and the people involved remember it. 
\end{abstract}
\section{Introduction}
Contracts are the most fundamental part of the law. A contract defines the obligations and rights of parties, who agree to work together according to the contract~\cite{schwartz2003contract,hermalin2007contract,chen2012contract}. A contract is a binding agreement between two or more parties. It is an offer by a party to do or not do something in exchange for a return by the other party. In the era of information technology, when the leak of confidential information can have devastating consequences, there is a need to keep contracts confidential. The trade-off of secrecy is the difficulty of verification, which we refer to in this paper as secrecy-verifiability-paradox (SVP).

There are many reasons why a contract might need to be kept private.  In some cases, there may be information in the contract that is sensitive or proprietary, and the parties may want to keep it confidential to prevent it from being disclosed to competitors. However, this secrecy comes at a cost; when a legal dispute occurs, no one can be sure that the presented data is authentic. Without a trusted third party, it is difficult to prove that the data has not been tampered with. In the traditional contract, notarization is used to provide this assurance. However, third parties are not used in the context of smart contracts. As a result, there need to be new mechanisms for proving the authenticity of data.

In another hand, blockchain is designed to  provide a decentralized,  secure and tamper-resistant database. As blockchain technologies mature, there are increasing opportunities for them to be used in new applications \cite{fernandez2019review,haleem2021blockchain,son2019blockchain,zheng2017overview,lao2020survey}. It is a chain of blocks that contains a cryptographic hash of the previous block, each block contains the hash of the last block, and the chain is tamper proof. In classic blockchains, such as Bitcoin, data is stored in the form of transactions. When information is  stored in the blockchain, it is immutable, which makes it tamper proof. When a transaction is created, the information is hashed and stored in the blockchain. If a transaction is changed, the hash will change, which makes it easier to detect the transaction is tampered with. 

The decentralization  provided by blockchain  is an important advantage over traditional centralized databases. In the traditional database, the information is stored at a central location and stored by an entity that is responsible for that information. However, in a decentralized database, the information is stored in multiple locations, and each entity is responsible for their own data. The decentralization  of the database made it more secure, because it is difficult to tamper with information stored in multiple locations. 
However, this nature is also the threat of the secrecy of the contract. Many  people have a copy of the information stored in the blockchain. As a result, if the contract is stored in the plain text, it is impossible to keep the contract secret. 

In this paper, we propose a new blockchain-based mechanism to solve the SVP problem. In our proposal, the contracts are stored as encrypted evidence in the blockchain, and the content can be protected from disclosure. However, when a dispute occurs, there are effective means to verify the content. Our proposed method allows different levels of revealment of details of the contract. With zero-knowledge proof, the arbitrator can verify that the contract is valid without leaking any information about the contract. 


\section{Preliminary}
Cryptography \cite{douglas1995cryptography} is a mathematical science that enables the secure transfer of information. The most acient form of cryptography is the Caesar Cipher \cite{goyal2013modified}, which was used by Julius Caesar to securely transfer messages to his military commanders during military campaigns. The method is simple as the alphabet is shifted by a fixed number, usually 3, and it is decrypted by shifting the alphabet back by the same amount. The general form of the Caesar Cipher is the shift cipher, where the alphabet is shifted by a fixed amount. For example, the message ``This is plaintext'' would be encrypted to ``Wklv lv sodlqwhaw'' by shifting each character by 3.

Cryptography has advanced significantly over the past two millennia from the Caesar Cipher to the RSA-Cryptosystem \cite{milanov2009rsa} which is one of the most widely used cryptosystems in the world. The RSA cryptosystem is a public-key cryptosystem and it is widely used for secure data transfer. It is based on the fact that it is easy to multiply two large prime numbers, but very difficult to factor a large number that is the product of two primes.
The RSA cryptosystem relies heavily on the number theory theorem that states that there are infinitely many prime numbers. The RSA cryptosystem is used for both encryption and decryption. The sender uses the recipient's public key to encrypt the message and the recipient uses their private key to decrypt the message. 

The SHA-256 algorithm \cite{gilbert2003security} is a cryptographic hash function that is used to generate a 256-bit hash value from an arbitrary length message. The cryptographic hash function has the property that it is very difficult to generate a message that has a given hash value. The SHA-256 algorithm is used in the Bitcoin protocol \cite{nakamoto2008bitcoin} to generate a unique address for each user.
A cryptographic hash function is a mathematical function that converts some input data into a hash code. 
This function is one-way, meaning that given the data it is very easy to calculate the hash code 
but given the hash code, it is very difficult to calculate the data. 
In addition, the cryptographic hash function is collision-resistant, meaning that it is very difficult to find two pieces of data that will result in the same hash code. 

Blockchain is distributed ledger technology that can be used to implement cryptocurrencies, such as Bitcoin and Ethereum \cite{dannen2017introducing}. 
A distributed ledger is a database that is maintained and synchronized by a network of computers that can be accessed by anyone with the proper permissions. 
This is different from a traditional database that is maintained by a single entity. 
A blockchain is a type of distributed ledger that uses cryptographic hashing to implement a chain of blocks, where each block contains a timestamp and a link to the previous block. 
This chain of blocks is distributed to all the computers in the network, and each computer verifies the chain of blocks to ensure that it is valid. 
Blockchain is transparent and immutable, meaning that anyone can view the contents of the blockchain, 
and once data is written to the blockchain, it cannot be modified. 
This makes blockchain an attractive option for implementing cryptocurrencies, 
since it provides a secure and tamper-proof way to track ownership of digital assets.
For smart contracts, blockchain allows for the execution of code in a decentralized manner, 
which can be used to implement various applications, such as decentralized exchanges and decentralized autonomous organizations \cite{buterin2014next}. 

Zero-knowledge-proof \cite{goldwasser1985knowledge} (ZKP) is a proof system where one party (the prover) can prove to another party (the verifier) that they know a value $x$, without conveying any information apart from the fact that they know the value $x$.
As a simple example, suppose there is a gold number between a range of numbers $[1, 100]$. The prover can prove that they know the gold number without conveying what the gold number is. The verifier is armed with a black box that can answer only one query: "Is there a gold number between $a$ and $b$?".
The verifier can verify that prover actually knows what is the gold number by asking a series of questions where the answer is either ``yes" or ``no".
This example is just to demonstrate the idea because, with enough queries, the verifier will eventually know what the gold number is.
Zero-knowledge-proof algorithms based on cryptographic theories are much more sophisticated and can prove the existence of an element without revealing any information about it.

\section{Secrecy-Verifiability Paradox and Proposed Solution}
Imagine that we want to write a contract that specifies a set of rules that we and other parties will follow, but we don't want to reveal the details of the contract to anyone else. At the same time, we want the contract's existence and terms to be verifiable.
This problem can not be solved by traditional means. Even that the current existing blockchain technology can not provide a perfect solution to this problem, as the details of the contract would be revealed if the contract is written on the blockchain. However, we propose a way to achieve both secrecy and verifiability by using a combination of blockchain technology and zero-knowledge-proof. 

\begin{definition}
A \emph{Zero-knowledge-proof of Knowledge} (ZKPOK) is a triple $(P,V,d)$ consisting of:

\begin{itemize}
    \item Prover $P$
    \item Verifier $V$
    \item A \emph{description} $d$ of the verifier's algorithm, consisting of:
    \begin{itemize}
        \item A finite set of statements $S$
        \item A finite set of questions $Q$
        \item A \emph{knowledge predicate} $K(x, y)$ over $S \times Q$
    \end{itemize}
\end{itemize}

\noindent
such that the following conditions hold:

\begin{itemize}
    \item If $P$ is given $x \in S$, $y \in Q$, and $K(x, y) = 1$ then $P$ can produce a \emph{proof} $p$ such that $V(x, y, p) = 1$.
    \item If $P$ is given $x \in S$, $y \in Q$, and $K(x, y) = 0$ then for all $p$, $V(x, y, p) = 0$.
\end{itemize}
\end{definition}

\begin{figure}
	\centering
	\includegraphics[width=.8\linewidth]{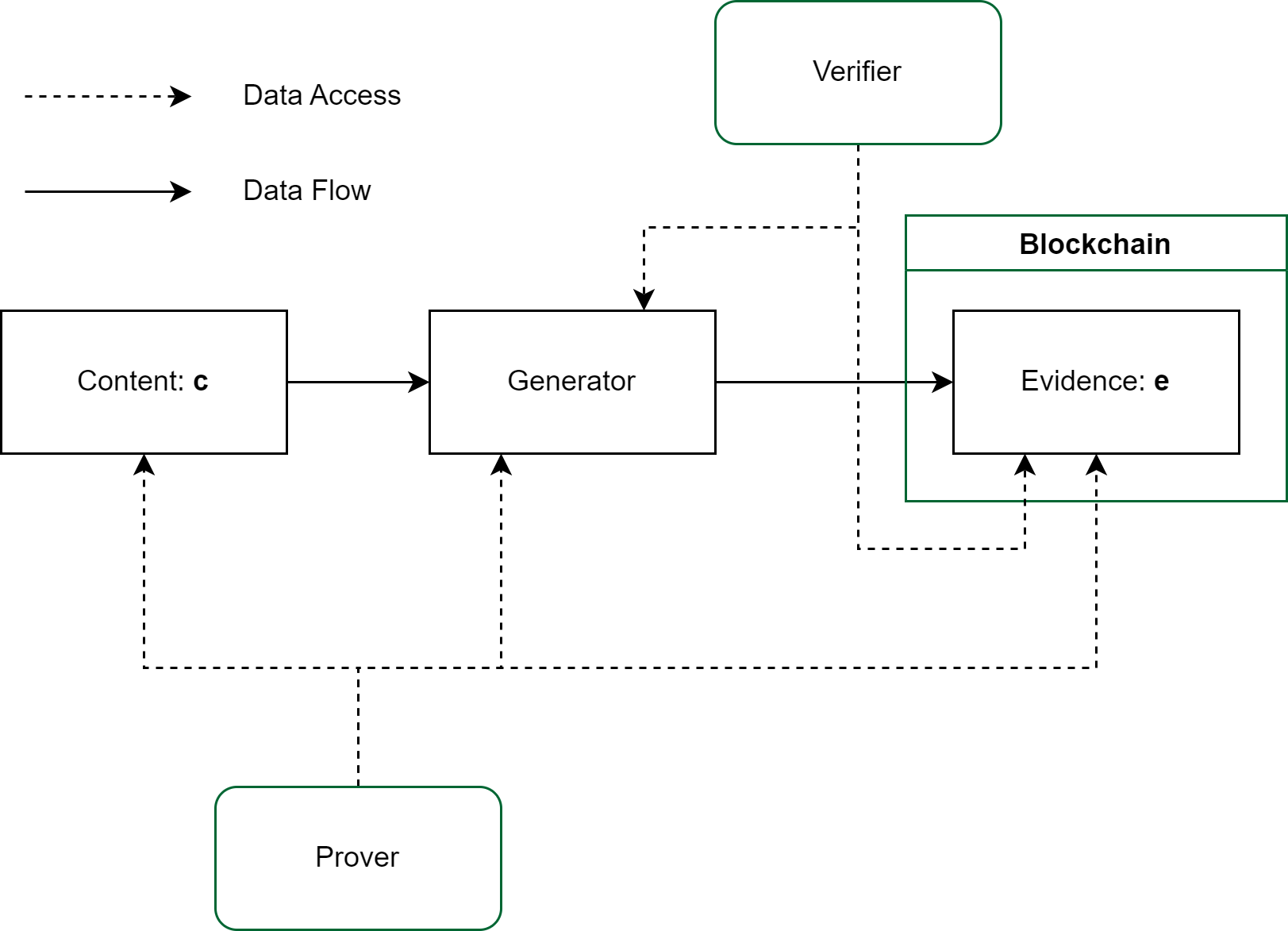}
	\caption{Overview of our proposed method.}
	\label{fig:diagram}
\end{figure}

Solving the problem of SVP, the contract \emph{content} $c$ should be kept secret, while the \emph{evidence} $e$  value should be published.
Figure \ref{fig:diagram} demonstrates the idea of our proposed method. The evidence is calculated from the content by a generator and stored on the blockchain. The prover and verifier access different information so that the contract content is not revealed but can be verified via the evidence.

For example, suppose that $e = g^{x} \bmod p$, one possible way to implement the verification process is as follows:

\begin{enumerate}
    \item $P$ chooses a random $r \in \mathbb{Z}_p$.
    \item $P$ computes $s = g^r \bmod p$ and sends $s$ to $V$.
    \item $V$ chooses a random $i \in \{0, 1\}$ and sends it to $P$.
    \item If $i = 0$ then $P$ sends $r$ to $V$.
    \item If $i = 1$ then $P$ computes $r' = r + c \bmod p$ and sends $r'$ to $V$.
    \item $V$ checks that $s = g^{r'} \bmod p$ and accepts if this is the case, otherwise rejects.
\end{enumerate}

The verifier can use $e$ value to calculate $s$ and compare it with the value received from the prover and does not learn anything about the contract $c$.
The above process can be done in multiple rounds to reduce the probability of cheating. In each round, the verifier sends a random value to the prover, and the prover responds with a value that depends on the received value and the secret value $c$. After a sufficient number of rounds, the verifier can be convinced with high probability that the prover knows the value $c$ without learning anything about $c$.

We can easily prove that the proposed method is sound, complete, and zero-knowledge.

\begin{proposition}
The above protocol is sound, i.e. it is impossible for a cheating prover to convince the verifier unless he knows the value of $c$.
\end{proposition}

\begin{proof}
Assume that a cheating prover $P$ does not know the value of $c$, but he wants to convince the verifier $V$ that he knows $c$. Since $i$ value is chosen randomly by $V$, in order for $P$ to cheat, $P$ needs to be able to predict the value of $i$. The probability of knowing the value of $i$ is $1/2$, and this means that the probability of cheating is $1/2$ as well. With multiple rounds, the probability of cheating becomes arbitrarily small.
\end{proof}

\begin{proposition}
The above protocol is complete, i.e. if $P$ knows the value of $c$ then he can convince $V$ that he knows $c$.
\end{proposition}

\begin{proof}
Given that $P$ knows the value of $c$, he can respond correctly to the questions of $V$ and convince $V$ that he knows $c$.
\end{proof}

\begin{proposition}
The above protocol is zero-knowledge, i.e. the value of $c$ is not revealed to $V$ even after $P$ convinced $V$ that he knows $c$.
\end{proposition}

\begin{proof}
As we can see, the value of $c$ is not revealed to $V$ and $V$ can not calculate $c$ from the received responses of $P$.
\end{proof}

\section{Discussions}

\subsection{Privacy}
Although the verifier does not know the contract content, the prover knows it. In this case, the prover is the \emph{contract initiator}, and it is assumed that the initiator can be trusted. We can use the aforementioned ZKPOK protocol to prove to the verifier that the initiator knows the contract content. The initiator can also use the ZKPOK protocol to prove to the other parties that he knows the contract content. Since the contract content is not stored on a blockchain, it can only be verified by the parties involved in the contract. 

\subsection{Solution Robustness}

As we can see, the contract content is private. The contract terms are not designed to be stored on the blockchain and are only known to the parties involved in the contract. Since the contract terms are not stored on the blockchain, only the parties involved can know and prove the existence of the contract. If a party loses the original contract,  they can ask other parties to prove the existence of their contracts. Furthermore, if this solution is implemented well, the party involved can prove the contract's existence with only a few essential unforgettable details. For example, in a lodging contract,  the party can prove the existence of the agreement with information about the period and the address of the property. 
\subsection{Security}

The security of the proposed method is the same as the security of the traditional cryptographic methods. In order to achieve higher security, the number of rounds can be increased. In addition to randomness, we can also ask the prover to prove that he knows the values of different terms in different rounds. Applying the divide-and-conquer technique, we can be convinced that the prover knows the contract content in polynomial time. Since the evidence is stored on a blockchain, we can verify the contract even in the case that its existence has been destroyed as long as the contract is real and the people involved remember it.

\subsection{Contract termination and modification}

The proposed method is not designed to verify the contract termination and modification. These terms should be verified by the parties involved in the contract. For the best protection, the contract should be terminated and modified with the consent of all parties. This consent can be written on the contract itself or in another contract.
The problem of inter-contractual relationships is beyond the scope of this paper but can be addressed in the future.

\section{Conclusion}

In this paper, we first point out the existence of a secrecy-verifiability paradox in smart contracts.
Without a trusted third party for notarization, it is impossible to prove the validity of a smart contract in a public ledger.
This paradox might be a critical drawback of smart contracts.
We then propose a solution to the problem by using a combination of blockchain technology and zero-knowledge proof. 
The proposed method allows us to write a contract that can be protected from revelation but remain verifiable. 
We can easily prove that the proposed method is sound, complete, and zero-knowledge.
The proposed solution is general enough to be used for a wide range of applications and a variety of blockchain platforms with appropriate modifications. 
In future work, we plan to expand and investigate the problem of multiple-contract SVP.

\bibliographystyle{splncs04}
\bibliography{ref.bib}

\begin{thebibliography}{10}
\providecommand{\url}[1]{\texttt{#1}}
\providecommand{\urlprefix}{URL }
\providecommand{\doi}[1]{https://doi.org/#1}

\bibitem{buterin2014next}
Buterin, V., et~al.: A next-generation smart contract and decentralized
  application platform. white paper  \textbf{3}(37), ~2--1 (2014)

\bibitem{chen2012contract}
Chen-Wishart, M.: Contract law. Oxford University Press (2012)

\bibitem{dannen2017introducing}
Dannen, C.: Introducing Ethereum and solidity, vol.~1. Springer (2017)

\bibitem{douglas1995cryptography}
Douglas, R.S., et~al.: Cryptography theory and practice. CRC Press  (1995)

\bibitem{fernandez2019review}
Fernandez-Carames, T.M., Fraga-Lamas, P.: A review on the application of
  blockchain to the next generation of cybersecure industry 4.0 smart
  factories. Ieee Access  \textbf{7},  45201--45218 (2019)

\bibitem{gilbert2003security}
Gilbert, H., Handschuh, H.: Security analysis of sha-256 and sisters. In:
  International workshop on selected areas in cryptography. pp. 175--193.
  Springer (2003)

\bibitem{goldwasser1985knowledge}
GOLDWASSER, S.: The knowledge complexity of interactive proof systems. In:
  Proceedings of 17th ACM Symposium on Theory of Computing. pp. 291--304 (1985)

\bibitem{goyal2013modified}
Goyal, K., Kinger, S.: Modified caesar cipher for better security enhancement.
  International Journal of Computer Applications  \textbf{73}(3),  0975--8887
  (2013)

\bibitem{haleem2021blockchain}
Haleem, A., Javaid, M., Singh, R.P., Suman, R., Rab, S.: Blockchain technology
  applications in healthcare: An overview. International Journal of Intelligent
  Networks  \textbf{2},  130--139 (2021)

\bibitem{hermalin2007contract}
Hermalin, B.E., Katz, A.W., Craswell, R.: Contract law. Handbook of law and
  economics  \textbf{1},  3--138 (2007)

\bibitem{lao2020survey}
Lao, L., Li, Z., Hou, S., Xiao, B., Guo, S., Yang, Y.: A survey of iot
  applications in blockchain systems: Architecture, consensus, and traffic
  modeling. ACM Computing Surveys (CSUR)  \textbf{53}(1),  1--32 (2020)

\bibitem{milanov2009rsa}
Milanov, E.: The rsa algorithm. RSA laboratories pp. 1--11 (2009)

\bibitem{nakamoto2008bitcoin}
Nakamoto, S.: Bitcoin: A peer-to-peer electronic cash system. Decentralized
  Business Review p. 21260 (2008)

\bibitem{schwartz2003contract}
Schwartz, A., Scott, R.E.: Contract theory and the limits of contract law. Yale
  LJ  \textbf{113}, ~541 (2003)

\bibitem{son2019blockchain}
Son, K.T., Thang, N.T., Dong, T.M., Thanh, N.H.: Blockchain technology for data
  entirety. Sci Research  \textbf{6}(6), ~68 (2019)

\bibitem{zheng2017overview}
Zheng, Z., Xie, S., Dai, H., Chen, X., Wang, H.: An overview of blockchain
  technology: Architecture, consensus, and future trends. In: 2017 IEEE
  international congress on big data (BigData congress). pp. 557--564. Ieee
  (2017)

\end{thebibliography}
\end{document}